\documentclass[preprint,aps,prc,showpacs,tightenlines,endfloats]{revtex4}
\usepackage{graphicx,amsmath,amsfonts}

\begin{document}

\preprint{ }

\title{Effects of negative energy components \\
in two-body deuteron photodisintegration}
\author{\firstname{Konstantin} Yu. \surname{Kazakov}}
\email{kazakovk@ifit.phys.dvgu.ru}
\author{\firstname{Denis} V. \surname{Shulga}}
\email{denis@ifit.phys.dvgu.ru} \affiliation{ Laboratory of
Theoretical Nuclear Physics, Far Eastern State University,
Sukhanov Str. 8, Vladivostok, 690600, Russia}
\date{\today}

\begin{abstract}

Several observables in two-body deuteron photodisintegration are
investigated in the framework of the Bethe-Salpeter formalism.
Apart from keeping throughout Lorentz covariance, a special
attention is paid to inclusion of both the positive energy and
negative energy partial-wave components of the deuteron state.
Using the  Bethe-Salpeter equation for deuteron in the ladder
approximation with one-boson exchanges as a driving force, the
contribution of the negative energy states is studied for the
unpolarized differential cross as well as the linear photon and
tensor target asymmetries. These states are found to have an
impact on the observables and, thus, should be taken into account
in a complete theoretical development of the reaction in the
intermediate energy regime.

\end{abstract}

\pacs{25.10+s; 25.20.Dc; 21.45+v; 11.10.St}

\maketitle

\section{\label{sec:intro}Introduction}

An application of the Bethe-Salpeter equation (BSE) for spinors
particles and Mandelstam's theory to the analysis of two-body
deuteron photodisintegration has been elaborated in the previous
works~\cite{mypaper1,mypaper2}, where the contributions of
relativistic effects to the differential cross section have been
estimated by using the Bethe-Salpeter (BS) formalism as well as
the equal-time approximation to the BSE. The analysis shows some
deficiency of without taking into account the full Dirac structure
of the two-nucleon bound state. It is well known that while
dealing with the BSE the problem of many coupled states arises
since there are four types of solutions of the Dirac equation for
a given momentum, two positive energy states and two negative
energy states. In our case this deficiency is ascribed to the
making use of the BSE in the context of the relativistic separable
interaction kernel with leaving out various negative energy
states~\cite{graz-tjon}. The present paper is to shed a light on
the effect of the inclusion of negative energy components of the
deuteron state on the observables of deuteron photodisintegration.
Their driving force is the one-boson exchange (OBE) kernel, in
which the pion-nucleon coupling is described with the help of
axial-vector (A) theory.

The relativistic covariant BSE with a superposition of $\pi$,
$\omega$, $\rho$, $\eta$, and $\delta$ exchanges originally has
been applied to a description of low energy NN scattering (see
Ref.~\cite{NN-tjon} and references therein). A good agreement with
the experimental data is shown  to be achieved for partial waves
with $J>0$ considering that the coupling for pion-nucleon vertex
to be an A type, i.~e. a weak $N\Bar N\pi$ coupling. Then the same
model has been implemented to a research on the deuteron
electromagnetic (EM) form factors in the relativistic
framework~\cite{EM-tjon}. These investigations have inspired
different authors to carry out studies of the EM properties of two
nucleon system~\cite{honzawa} and development of an effective
theory of strongly interacting particles at momentum transfers of
a few GeV/c~\cite{gross}. Moreover, a deuteron resulting from this
model has been obtained from the solution of the BSE in the ladder
approximation with basic mesons and by adding one more meson, the
$\sigma$~\cite{umnikov-khanna}. Based on this numerical solution,
extended calculations of the relativistic corrections to the
deuteron static properties, such as the magnetic and quadrupole
moments of deuteron, have been made~\cite{kapkaz}.

The work introduced by this paper is an realization of this
relativistic framework for studying two-body deuteron
photodisintegration. As energy and momentum transfers involved in
the process are held to be below one GeV, the appropriate
effective degrees of freedom in this energy range are the mesons
and nucleons. Working in the plane-wave one-body approximation, we
give an accurate and thorough treatment of the differential cross
section, the photon beam well as three tensor asymmetries. The
effects of various relativistic contributions to these observables
are examined using the BSE with two rather different kernels: the
OBE and separable ones. Our ultimate objective is to establish a
role played by the negative energy states of the relativistic wave
function of deuteron.

The next section briefly refers to the approach used. Definitions
of the polarization observables, the linear photon asymmetry
$\Sigma^l$ and tensor-target asymmetries $T_{2M}$ ($M=0,1,2$) are
given in terms of reaction amplitudes. Since the chief source of
the relativistic effects comes from the relativistic wave function
of deuteron, we discuss decomposition of the full amplitude into
partial waves. The deuteron properties in the framework of the BS
formalism are also specified. Sec.~\ref{sec:amplitude} shows the
general structure of the reaction amplitudes in the plane-wave
case with the one-body EM current contribution. In particular, the
inclusion of both the positive and negative energy states of the
deuteron vertex function as well as boost operator is considered.
In Sec.~\ref{sec:results} we give numerical results for the
observables with an emphasis on the leading contribution coming
from the dominant negative-energy triplet $P$-component for the
OBE kernel in A theory. Due to some core reasons, our numerical
work is restricted to a case when one keeps the relative energy
variable of the deuteron vertex function equal to zero. A
comparison with the previous calculations, performed for the
separable kernel, is also carried out and differences are
discussed. Sec.~\ref{sec:remarks} concludes the paper with a
number of authors' remarks on the subject.

\section{Review of the model}
\label{sec:review}

We briefly describe the basics of the relativistic model
presented in Refs.~\cite{mypaper1,mypaper2}.

\subsection{Definition of the polarization observables}

The description of all the possible polarization observables in
the reaction $\gamma+D\to P+N$ with polarized photons and oriented
deuterons has been given in Ref.~\cite{arenh}. The treatment uses
a standard coordinate system with $z$-axis, chosen as quantization
one, in the direction of the incoming photon c.~m. three momentum
$\mathbf{q}$ ($|\mathbf q|\equiv\omega$) and $x$ axis in the
direction of maximal linear polarization of the photon. Spin
indices of the initial state are specified by the photon
polarization $\lambda=\pm1$ and the deuteron spin projection
$m_d=0,\pm1$ with respect to the quantization axis. The final
state is given by a free neutron-proton ($np$) with the relative
three momentum $\mathbf{p}$ and by the total spin $S=0,1$ and its
$z$-projection $m_s$. Concerning only angular dependence, the
reduced reaction $t$ matrix elements are expressed in terms of
the reaction $T$ matrix for two-body photodisintegration of
deuteron in the c.m. frame as follows
\begin{equation}\label{}
t_{Sm_s\,\lambda
m_d}(\Theta_p)=\text{e}^{-\imath(\lambda+m_d)\Phi_p}
T_{Sm_s\,\lambda m_d}(\Phi_p,\Theta_p),
\end{equation}
where $\Theta_p$ and $\Phi_p$ are the spherical angles of the
$np$ pair relative three momentum with respect to the frame of
reference.

Before the form of $T_{Sm_s\,\lambda m_d}$  is specified in the
framework of the BS theory and approximation is discussed, we
write the most general expressions for the polarization
observables in question as expressed of products of the reaction
amplitudes. First, it is the unpolarized differential cross
section
\begin{align}
\dfrac{d\sigma_0}{d\Omega_p} &=\dfrac{\alpha}{16\pi
s}\dfrac{|\mathbf p|}{\omega}F(\Theta_p),\qquad
F(\Theta_p)=\dfrac13\sum\limits_{Sm_S m_d} |t_{Sm_s\,\lambda=1
m_d}(\Theta_p)|^2,\label{diff}
\end{align}
where $\alpha=e^2/(4\pi)$ is fine structure constant and $s$ is
the square of the total energy of the $np$ pair. Second, the
photon asymmetry for linearly polarized photons and the tensor
target asymmetries
\begin{align}
-\Sigma^l F(\Theta_p) &=\biggl.\dfrac13\sum\limits_{S m_Sm_d}
t_{Sm_s\,1
m_d}^*t_{Sm_s\,-1 m_d}, \label{asymm}\\
T_{2M}F(\Theta_p) &=\dfrac{\sqrt{5}}{3} \sum\limits_{S
m_Sm_d}\text{C}_{1m_d 2M}^{1M+m_d}\:\text{Re}\bigr\{{t_{Sm_s\,1
m_d}^*t_{Sm_s\,1 M+m_d}}\bigl\}(2-\delta_{M0}),\qquad
(M\geq0)\label{t2m}.
\end{align}
In the Eqs.~\eqref{diff}-\eqref{asymm} it is assumed that te
observables also depend on the photon lab. energy $E_\gamma$, for
which defines
\begin{equation}
\omega=\frac{M_d}{\sqrt{s}}E_\gamma\quad\text{with}\quad
s=M_d(M_d+2E_\gamma)\label{kin-s},
\end{equation}
where $M_d$ is the deuteron rest mass.

The reaction $T$ matrix is expressed in terms of the matrix
elements of the EM current between the final and initial 2N
states. The procedure for the calculation of the matrix elements
in the BS theory is based on Mandelstam's theory and the reduction
formalism. It also preserves the consistence between the
amplitudes and current operators from the very outset, see the
example in Refs.~\cite{benz,shebeko}. In the deuteron breakup one
deals with the matrix elements of the EM current between two-body
bound state (deuteron) and asymptotically free $np$~scattering
state
\begin{equation}\label{T-amp}
T_{Sm_s\,\lambda m_d}=\dfrac{1}{4\pi^3}
 \int\!\!d^4kd^4u \: \overline{\chi}_{Sm_s}
(u;\Hat p\,\mathbb P)\,\epsilon_\lambda\!\cdot\! J(u,k;q,\mathbb
K)\chi_{m_d}(k;\mathbb K),
\end{equation}
where $J$ is the irreducible EM vertex, $\chi_{m_d}(k;\mathbb K)$
is the BS amplitudes for deuteron with the total momentum $\mathbb
K=(E_d,-\boldsymbol{\omega})$ with
$E_d=\sqrt{M_d^2+\boldsymbol{\omega}^2}$ and
$\overline{\chi}_{Sm_s}(k;\Hat p\,\mathbb P)$ denotes the
conjugate BS amplitude of $np$ pair with $\Hat p=(0,\mathbf p)$
and $\mathbb P=(\sqrt{s},\mathbf 0)$ being the on-mass-shell
relative and the total four momenta.

The EM vertex operator $J$ is split up into two pieces
\begin{equation}\label{}
J=J^{[1]}+J^{[2]},
\end{equation}
where $J^{[1]}$ is a free part (a nucleon couples individually to
the radiation field without interacting with another) and
$J^{[2]}$ is a two-body EM vertex operator. Owing to the facts
that the deuteron is an isoscalar $I=0$ target and EM interaction
does not conserve the total isospin $I$, the two-body vertex
correction contributes to the Eq.~\eqref{T-amp}. The gauge
independence of the reaction amplitude $q\cdot T=0$ is necessarily
guaranteed by the one- and two-body Ward-Takahashi identities,
which the EM vertexes $J^{[1,2]}$ have to satisfy. The sufficient
condition of the gauge independence demands both amplitudes to be
solutions of the BSE with the same irreducible interaction kernel
$\mathcal V$. In the ladder approximation the corresponding
equations read
\begin{align}
\chi_{m_d}(k;\mathbb K )&= \dfrac{\imath}{4\pi^3}\int\!\!d^4 u\,
S^{(1)}\Bigl(\dfrac{\mathbb K}{2}+u\Bigr)
S^{(2)}\Bigl(\dfrac{\mathbb K}{2}-u\Bigr)
\mathcal V(k,u)\chi_{m_d}(u;\mathbb K),\label{BS-hom}\\
\chi_{Sm_s}(k;\Hat p\,\mathbb P)&= \chi_{Sm_s}^{(0)}(k;\Hat
p\,\mathbb P)+ \dfrac{\imath}{4\pi^3}\int\!\!d^4 u\,
S^{(1)}\Bigl(\dfrac{\mathbb K}{2}+u\Bigr)
S^{(2)}\Bigl(\dfrac{\mathbb K}{2}-u\Bigr)\mathcal
V(k,u)\chi_{Sm_s}(u;\Hat p\,\mathbb P),\label{BS-inhom}
\end{align}
where $S^{(l)}$ is the free nucleon propagator $(l=1,2)$ and
$\chi_{Sm_s}^{(0)}(k;\Hat p\,\mathbb P)$ stands for the amplitude
for the motion of free particles.

\subsection{The bound state vertex function}

Now we outline ideas concerning the partial-wave decomposition of
the relativistic amplitude of deuteron. One needs the vertex
function in order to calculate the EM current matrix elements. Its
form can be readily determined in a general moving frame, after it
is obtained in the deuteron rest frame.

The vertex function is related to the BS amplitude through the
relation
\begin{equation}\label{deuteron-vertex}
\Gamma_{m_d}(k;\mathbb K)=\left[S^{(1)}\Bigl(\dfrac{\mathbb
K}{2}+k\Bigr)S^{(2)}\Bigl(\dfrac{\mathbb
K}{2}-k\Bigr)\right]^{-1}\chi_{m_d}(k;\mathbb K).
\end{equation}
There are two ways to describe the partial-wave decomposition of
the vertex function in the Eq.~\eqref{deuteron-vertex}: the direct
product and matrix ones. The latter allows to nicely absorb the
angular-dependence factors into the specification of the
partial-wave states entirely. Relevant explanations can be found
in Ref.~\cite{bs-matrix}. In both cases the general forms of the
eigenstates of the BSE for a given total angular momentum $J$ are
classified according to the spatial parity and "exchange" quantum
number that embodies the Pauli principle for two identical
relativistic particles~\cite{kubis}. For the coupled physical
channels $^3S_1$--${}^3D_1$ the relativistic wave function
consists of eight states. In the momentum space, the deuteron
vertex function is given by
\begin{equation}\label{}
\Gamma_{m_d}(k_0,\mathbf k;\mathbb
K_{(0)})=\sum\limits_{\alpha=1}^8 g(k_0,|\mathbf
k|,\alpha)\Gamma_{m_d}(-\mathbf k,\alpha)\zeta_0^0,
\end{equation}
where $\mathbb K_{(0)}=(M_d,\mathbf 0)$, $\Gamma_{m_d}(-\mathbf
k,\alpha)$ are the normalized spin-angular momentum
eigenfunctions involving the Dirac $u$ and $v$ spinors,
$\zeta^{I_3}_I$ denotes the normalized eigenstate of the total
isospin $I$ and $I_3$. Index labeled as
$\alpha={}^{2S+1}L_{J=1}^{\rho_1\rho_2}$ runs over all
symmetrical under interchanged of particle 2N states with the
total orbital momentum $L$=0--2, the spin $S=0,1$ and the energy
spins of both particles $\rho_l=\pm 1$:
\begin{equation}\label{alpha-states}
\begin{split}
&1:{}^3S_1^{++},\quad 2:{}^3D_1^{++},\quad3:{}^3S_1^{--},\quad
4:{}^3D_1^{--},\\
&5:{}^3P_1^{+-},\quad6:{}^3P_1^{-+},\quad7:{}^1P_1^{+-},\quad8:{}^1P_1^{-+}.
\end{split}
\end{equation}
The first two states have the nonrelativistic counterparts. The
rest six states, corresponding to the negative energy one-particle
states, are the relativistic components of the deuteron vertex
function.

The partial-wave decomposition of the BSE~\eqref{bs-eqn} over
states in Eq.~\eqref{alpha-states} yields the set of the coupled
two-fold integral equations in the relative energy $k_0$ and
modulo of the relative three-momentum $|\mathbf k|$
\begin{equation}\label{bs-eqn}
g(k_0,|\mathbf k|,\alpha)= \dfrac{\imath}{4\pi^3} \int\!\! dq_0
d|\mathbf q|\sum\limits_{\beta,\gamma}\mathcal V(k_0,|\mathbf
k|,\alpha;q_0,|\mathbf q|,\beta)G_{0}(q_0,|\mathbf
q|,\beta,\gamma)g(q_0,|\mathbf q|,\gamma),
\end{equation}
where $\mathcal V(k_0,|\mathbf k|,\alpha;q_0,|\mathbf q|,\beta)$
and $G_{0}(q_0,|\mathbf q|,\beta,\gamma)$ are the partial-wave
projections of the interaction kernel $\mathcal V$ and the
two-particle spinor propagator $G_0=S^{(1)}S^{(2)}$ onto the basis
two-nucleon states~\eqref{alpha-states}. The coupling between
positive and negative energy states occurs directly through the
interaction kernel matrix and indirectly through the propagator
matrix. The term $G$ has a simple form independent of angle and
spin variables, as it depends only on the $\rho$-spin indices. The
corresponding structure of the matrix $\mathcal V(k_0,|\mathbf
k|,\beta;q_0,|\mathbf q|,\gamma)$ for pseudoscalar and scalar
exchanges can found in Ref.~\cite{kubis} and for axial-vector,
vector and tensor exchanges in Ref.~\cite{I1-tjon}. As for the
numerical part, there have been three solutions of the
BSE~\eqref{bs-eqn} up to now. Two of them are appropriate for the
interaction kernel which is a superposition of the one-boson
exchanges~\cite{EM-tjon, umnikov-khanna}. The third one is done
for the multi-rank separable interaction kernel, which is a
relativistic covariant generalization of the nonrelativistic
separable Graz-II potential for the NN system in the coupled
$^3S_1$ and $^3D_1$ states~\cite{graz-tjon}.

\begin{figure}[b]
\centering
\includegraphics[width=0.75\textwidth]{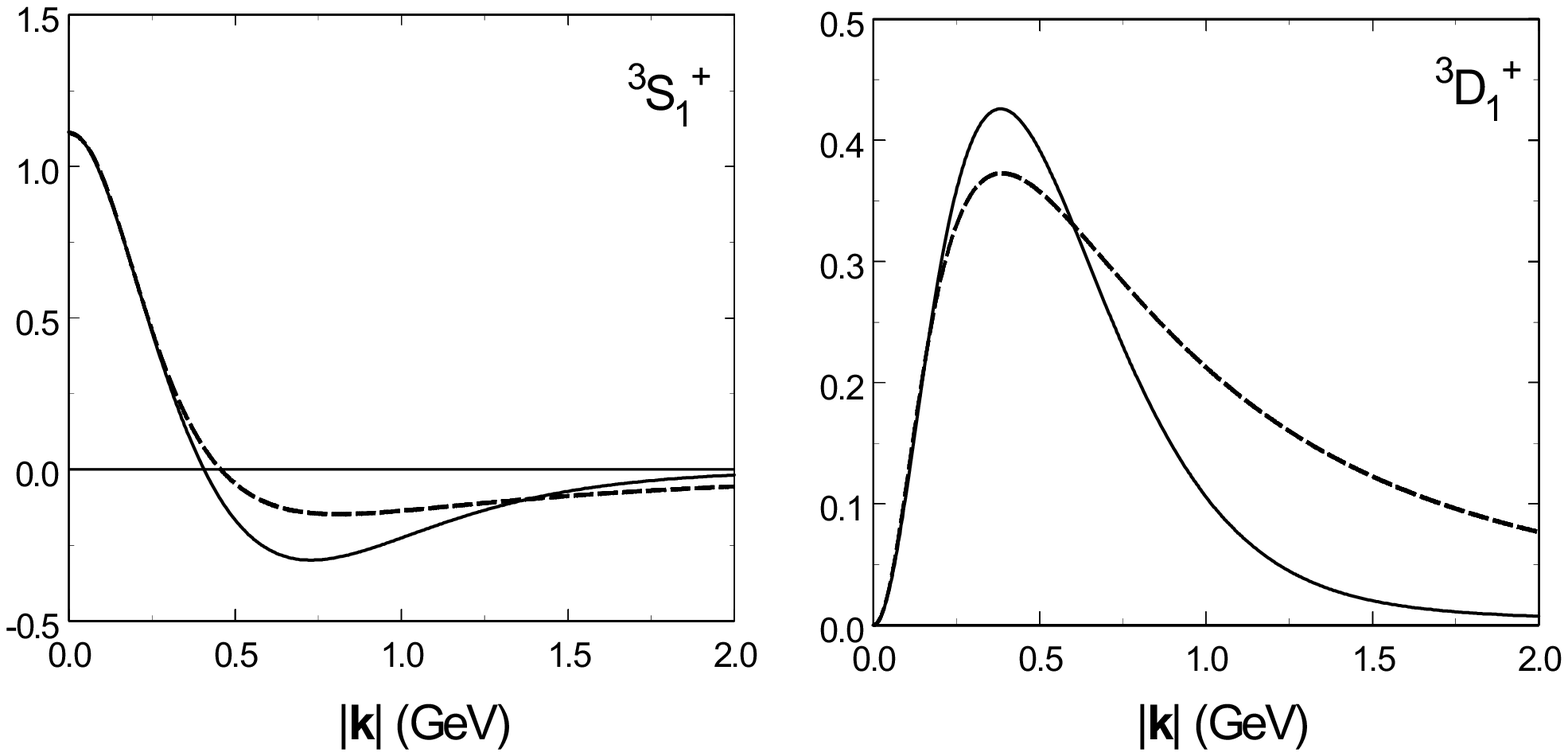}
\caption{\label{fig1} The momentum dependence of the radial
components of the deuteron vertex function
($^1S_1^{+}$--$^1S_1^{+}$ channel) is shown for the relative
energy variable $p_4=0$. Solid and dash curves depict the behavior
of the vertex functions for the one-boson exchange and
relativistic separable interactions, respectively.}
\end{figure}

In this paper we also present the research on the question how
strongly the observables of deuteron break-up are sensitive to the
different inputs of the deuteron vertex functions. Fig.~\ref{fig1}
shows the the positive energy components $g(k_0,|\mathbf
k|,\alpha)$ $(\alpha=1,2)$ for two classes of the interaction
plotted against the momentum $|\mathbf k|$ at $k_0=0$.
It is seen that the behavior in certain regions essentially
depends on the type of interaction model used for a description of
the NN system. The structures are similar at low three-momentum
$|\mathbf k|\leq 250$~MeV. A `tail` at high momenta of the
$^3D_1^{+}$-component for the separable interaction is much harder
than those for the OBE one\footnote{The factor $\imath^2$ is
absorbed in the definition of the $^3D_1^{+}$ partial-wave
component.}. The $^3S_1^+$-component for the OBE interaction has a
more pronounced dip with respect to that of the separable
interactions in the region $0.5\leq|\mathbf k|\leq 1$~GeV. These
differences have an perceptible effect upon the observables.

In further discussion it is worth introducing states, which
symmetrical and antisymmetrical with respect to the relative
energy $k_0$. These states are labelled by numbers
$n={}^{2S+1}L_1^\rho$ with $\rho=(+), (-), (\text{e})$ and
$(\text{o})$ being the projection of the total energy spin of the
2N system. In this case the notation of the components of the
vertex function is as follows,
\begin{equation}\label{n-states}
1:{}^3S_1^{+},\quad 2:{}^3D_1^{+},\quad3:{}^3S_1^{-},\quad
4:{}^3D_1^{-},\quad
5:{}^1P_1^{\text{e}},\quad6:{}^3P_1^{\text{o}},\quad7:{}^1P_1^{\text{o}},
\quad8:{}^3P_1^{\text{e}}.
\end{equation}
The first six states are even in the relative energy, while the
last two change signs as $k_0\to-k_0$, i.~e. they are referred to
as odd. The normalization condition for the vertex function is the
matrix element of the charge operator between the deuteron state
at the zero momentum  transfer
\begin{equation}\label{norm-cond}
\dfrac{1}{2\pi^2M_d}\sum\limits_{n=1}^8\int\limits_{-\infty}^{+\infty}\!\!
dk_4\int\limits_0^\infty\!\! d|\mathbf k| |\mathbf k|^2
\omega_{\rho}(E_k)\bigl[\phi(k_4,|\mathbf k|,n)\bigr]^2=1,
\end{equation}
where $k_4=\imath k_0$,
$\omega_{\rho}=\tfrac12(\rho_1+\rho_2)E_k-\dfrac{M_d}{2}$ and
$E_k=\sqrt{m^2+|\mathbf k|^2}$. The radial components of the
vertex function $g(n)$ are related to those $\phi(n)$ of the BS
amplitude as follows
\begin{align}
g(1,2)&=G_0(+,+)^{-1}\phi(1,2)\label{pp}\\
\intertext{for the $S$- and $D$-positive energy states,}
g(3,4)&=G_0(-,-)^{-1}\phi(3,4)\label{mm}\\
\intertext{for the $S$- and $D$-negative energy states,}
g(5,7)&=\dfrac{G_0(\text{e},\text{e})\phi(5,7)-G_0(\text{e},\text{o})\phi(6,8)}
{G_0(\text{e},\text{e})^2+G_0(\text{e},\text{o})^2}\label{even}\\
\intertext{for the $P$-states even in the relative energy,}
g(6,8)&=\dfrac{G_0(\text{e},\text{o})\phi(5,7)+G_0(\text{e},\text{e})\phi(6,8)}
{G_0(\text{e},\text{e})^2+G_0(\text{e},\text{o})^2}\label{odd}
\end{align}
for the $P$-states odd in the relative energy. In the
Eqs.~\eqref{pp}-\eqref{odd} the partial wave projections of the
two-nucleon propagator are labelled as a matrix in $\rho$-subspace
\begin{align}\label{}
G_{0}(+,+)&=\dfrac{1}{\bigl(\tfrac{M_d}{2}-E_k\bigr)^2+k_4^2},
&&G_{0}(-,-)=\dfrac{1}{\bigl(\tfrac{M_d}{2}+E_k\bigr)^2+k_4^2},\\
G_{0}(\text{e},\text{e})&=\dfrac{1}{2}[G_0(+,+)+G_0(-,-)],
&&G_{0}(\text{e},\text{o})=\dfrac{1}{2}[G_0(+,+)-G_0(-,-)].
\end{align}
The normalization condition~\eqref{norm-cond} defines the
$n$-state probability. According to Ref.~\cite{EM-tjon} each can
be understood as a measure of the effective charge of the state.
The corresponding numerical values are listed in
Table~\ref{tab:table1}.
\begin{table}[b]
\caption{\label{tab:table1} Deuteron properties in the framework
of the BS formalism, compared with experiment.}
\begin{ruledtabular}
\begin{tabular}{l l l l l}
& OBE & Graz-II & Empirical & Reference(s) \\ \hline Binding
energy $\varepsilon_d$~(MeV) & 2.2250 & 2.2250 & 2.224575(9) &
\cite{cd-bonn,kapkaz,graz-tjon}\\
Asymptotic $D/S$ ratio $\rho_{D/S}$ & 0.02497 & 0.02691 &
0.0256(4) & \cite{cd-bonn,umnikov-param,graz-tjon} \\
Quadrupole moment $Q_d$~(fm$^2$) & 0.2678\footnote{Without meson
current contributions and with relativistic corrections.}    &
0.2774\footnotemark[1]
& 0.2859(3) & \cite{cd-bonn,kapkaz,graz-tjon}\\
Magnetic moment $\mu_d$~$(\tfrac{e}{2m})$ &
0.8561\footnotemark[1] & 0.8512\footnotemark[1] & 0.857506(1) & \cite{cd-bonn,kapkaz,graz-tjon}\\
$D$-state probability $P_D$~(\%) & 5.10 & 5.0 &   & \cite{umnikov-param,graz-tjon}\\
Pseudoprobability $P_{-}$~(\%)   &  -0.0050 & --- & & \cite{umnikov-param}\\
Pseudoprobability $P_{\text{even}}$~(\%) & -0.0920 & --- & & \cite{umnikov-param}\\
Pseudoprobability $P_{\text{odd}}$~(\%) & -0.0230 & --- &  & \cite{umnikov-param}\\
\end{tabular}
\end{ruledtabular}
\end{table}
The upper limit of integration in Eq.~\eqref{norm-cond} is limited
to 3 GeV both for the relative energy and momentum. One can see
that the probabilities for the relativistic components of the
vertex function are negative, and they will be further referred to
as "pseudoprobabilities". As a consequence, inclusion of the
negative energy states reinforce the contribution to the charge of
the positive energy ones. Their resultant contribution is greater
than one. The smallness pseudoprobabilities of the negative energy
states is explained by the use of the axial-vector coupling for
the $\pi N$-vertex. The dominant relativistic component is the
triplet $^3P_1^{\text{o}}$-wave with $P_6=-0.08$~\%, and
contribution of the $^3S_1^{-}$- and $^3D_1^{-}$-states are
negligible. In the case of the $\pi N$ interaction due to
pseudoscalar coupling, values of the pseudoprobabilities of the
negative-energy partial states are expected to be
characteristically  different. In that case the qualitative
analysis shows that $^3S_1^{-}$-state may have the largest
pseudoprobability, while all other negative-energy states may have
been dismissed~\cite{honzawa}.

In Table~\ref{tab:table1} we also quote the numerical values for
the quadrupole $Q_d$ and the magnetic $\mu_d$ moments (without
meson current contributions) as well as the asymptotic $D/S$-state
ration $\rho_{D/S}$. The moments $Q_d$ and $\mu_d$ are given with
taking into account of the relativistic corrections. The
experimental values are cited according to the ones given in Table
XVII of Ref.~\cite{cd-bonn}. In case of the OBE interactions the
contribution of the negative energy components is considered. It
is purely of the relativistic nature and it is negative in sign,
thus reinforcing the discrepancy with the empirical value. As the
Graz-II interaction favors the $D$-state probability of 5~\%, we
have focused our attention on the separable vertex function
calculated for this case. The asymptotic $D/S$-state ration has
been calculated for each vertex function explicitly. In momentum
space, the ratio of the $^3D_1^{+}$ over the $^3S_1^{+}$ component
of the vertex function with both particles on the mass-shell is
defined by the formula~\cite{EM-tjon}
\begin{equation}\label{}
\rho_{D/S}=\dfrac{g(k_4,|\mathbf k|,2)}{g(k_4,|\mathbf
k|,1)}\Biggr|_{k_4=0,E_k=\tfrac{M_d}{2}}.
\end{equation}
Extrapolation of the radial vertex functions to the unphysical
value of $|\mathbf k|^2=\tfrac{M_d^2}{4}-m^2$ is achieved in two
ways: directly and by expansion of the functions in the Taylor
series up to terms of the third order.

It is interesting to compare the pseudopropabilities of the
relativistic components of the deuteron vertex function of
Refs.~\cite{EM-tjon} and~\cite{kapkaz}. Both functions are
calculated using the BSE in the ladder approximation with the same
superposition of the meson exchanges, except that in
Ref.~\cite{kapkaz} one more $\sigma$ meson is added. Difference in
the pseudoprobabilities of the spin singlet and triplet
$P$-components reaches one order of magnitude. Compare the total
strength of these states $P_-=-2.5\times10^{-2}$ of
Ref.~\cite{EM-tjon}, which is mostly due to the spin-singlet even
and odd $P$-states, with $P=-1.1\times10^{-1}$ of
Ref.~\cite{kapkaz}. For the latter value the major contribution is
due to the spin-triplet even $P$-states. The source of such
difference may may reside in the structure of the interaction
matrix $\mathcal V$ in Eq.~\eqref{bs-eqn}. For the axial-vector,
vector and tensor exchanges (opposite to the scalar one) there is
a direct coupling between the partial-wave states that are even in
relative energy and states that are odd. Consequently, effects as
turning off the $\tfrac{q^\mu q^\nu}{\mu_V^2}$ term in the vector
propagators or a tiny change of the $\pi N$-coupling in A theory
may influence on a quite sensitive changes in strengths of the
negative energy channel.

\section{\label{sec:amplitude}The reaction amplitude}

Having determined the vertex function in the rest frame of the
deuteron, we can calculate the reaction amplitudes~\eqref{T-amp}
and, as a result, obtain the differential cross
section~\eqref{diff}, photon and the target
asymmetries~\eqref{asymm} and~\eqref{t2m}.

In the plane-wave case with the one-body EM vertex one finds that
(the detail derivation is given in Ref.~\cite{mypaper1})
\begin{align}\label{PWOB}
T_{Sm_s\,\lambda m_d}&=\sum_{l=1,2}\Bar\chi_{Sm_s}(\mathbf
p)\zeta_s^+\Lambda(\mathcal
L)\Gamma_\lambda^{(l)}(q^2=0)S^{(l)}\bigl(\dfrac{\mathbb
K_{(0)}}{2}-(-1)^l k_l\bigr)\Gamma_{m_d}(k_l;\mathbb K_{(0)})\\
&-\sum_{l=1,2}(-1)^S\Bar\chi_{Sm_s}(-\mathbf
p)\zeta_v^+\Lambda(\mathcal
L)\Gamma_\lambda^{(l)}(q^2=0)S^{(l)}\bigl(\dfrac{\mathbb
K_{(0)}}{2}+(-1)^l k_l\bigr)\Gamma_{m_d}(-k_l;\mathbb
K_{(0)}),\notag
\end{align}
where the conjugate of the 2N continuum amplitude
$\Bar\chi^{(0)}=\gamma_0^{(1)}\gamma_0^{(2)}\chi^{(0)}$ is
expressed in terms of the two free Dirac positive energy spinors
\begin{equation}\label{np-amp}
\chi_{Sm_s}(\mathbf p)=\sum\limits_{\lambda_1\lambda_2}
C^{Sm_s}_{\tfrac12\lambda_1\tfrac12\lambda_2}
u_{\lambda_1}(\mathbf p)u_{\lambda_2}(-\mathbf p)
\end{equation}
and the two combinations of the isospin singlet and triplet
functions $\zeta_s=\zeta_0^0+\zeta^0_1$,
$\zeta_v=\zeta_0^0-\zeta^0_1$. The operator $\Lambda(\mathcal
L)=\Lambda^{(1)}(\mathcal L) \Lambda^{(2)}(\mathcal L)$ represents
the Lorentz transformation $\mathcal L$ on the Dirac subspace with
\begin{equation}\label{Lamda}
\Lambda^{(l)}(\mathcal L)=\Bigl(\dfrac{E_d+M_d}{2M_d}\Bigr)^{1/2}
\Bigl[1+\dfrac{\gamma_0\boldsymbol{\gamma}\cdot\mathbf
q}{E_d+M_d}\Bigr]^{(l)}
\end{equation}
The matrix $\mathcal L$, defined as $\mathbb K=\mathcal L\mathbb
K_{(0)}$, boosts the initial BS amplitude from the rest frame to
the c.~m. frame, in which the deuteron moves with a velocity
$\omega/M_d$.

In deuteron break-up we deal with the half off-mass-shell
photon-nucleon vertex $\Gamma^{(l)}_\lambda$, since the
knocked-out nucleon is taken as the physical one. Moreover,
$\Gamma^{(l)}_\lambda$ takes on the on-shell form at the real
photon point $q^2=0$ as a consequence of gauge
invariance~\cite{nagorny}
\begin{equation}\label{gammaNN}
\Gamma(q^2=0)=\epsilon_\lambda^\mu\gamma_\mu\dfrac{1+\tau_3}{2}+
\dfrac{\imath}{2m}\sigma_{\mu\nu}\epsilon_\lambda^\mu
q^\nu\dfrac{\kappa_s+\kappa_v\tau_3}{2},
\end{equation}
where $\kappa_s=\kappa_p+\kappa_n$ and
$\kappa_v=\kappa_p-\kappa_n$ with the anomalous part of the
proton (neutron) magnetic moments in units of the nuclear
magneton $1/(2m)$ denoted as $\kappa_{p(n)}$.

As the result of the boosting of the bound state wave function
along the negative $z$ axis, the relative four-momentum $k_l$ in
Eq.~\eqref{PWOB} is ``contracted". One can find
\begin{align}\label{kl}
{k_l}_0&=\dfrac{\omega}{M_d}p_z+(-1)^l\dfrac{\sqrt{s}}{2M_d}\:\omega\notag,\\
{k_l}_x&=p_x,\quad {k_l}_y=p_y,\\
{k_l}_z&=
\dfrac{\sqrt{s}-\omega}{M_d}p_z+(-1)^l\dfrac{\sqrt{s}}{2M_d}\:\omega\notag.
\end{align}

We proceed with introducing the matrix representation for each
partial wave of the deuteron vertex function~\eqref{alpha-states}
and for the amplitude of $np$ pair~\eqref{np-amp}. The reaction
amplitudes~\eqref{PWOB} related to $\Delta I=0$ transitions
interfere with those having $\Delta I=1$. Evaluating the isospin
part of matrix elements, the amplitudes $T_{Sm_s\:\lambda
m_d}^{a}$ for the isoscalar $a=1$ and isovector $a=2$ transitions
are cast into traces of $\gamma$-matrix terms. Due to the symmetry
consideration, the first two terms in Eq.~\eqref{PWOB} are
identical to the letter two. Further evaluation of amplitudes for
the given set of the polarization indices $\lambda, m_d$ and $m_s$
has been analytically carried out with the help of the formulae
manipulating language REDUCE~\cite{reduce}. The resulting
expressions have the form
\begin{equation}\label{t-final}
\begin{split}
t_{Sm_s\lambda m_d}^{a}&=
\sum\limits_{l=1,2}\sum\limits_{\alpha=1}^8
\bigl[\tilde\Gamma_{Sm_s\lambda m_d}^a(\omega,\mathbf
k_l;l,\alpha)S^{(l)}_{\rho_1}({k_l}_0,|\mathbf
k_l|)g({k_l}_0,|\mathbf
k_l|;\alpha)\\
&+\tilde\Gamma_{Sm_s\lambda m_d}^a(\omega,-\mathbf
k_l;l,\alpha)S^{(l)}_{\rho_2}({k_l}_0,|\mathbf
k_l|)g(-{k_l}_0,|\mathbf k_l|;\alpha)\bigr],
\end{split}
\end{equation}
where quantities $S^{(l)}_{\rho_i}$ are the positive
($\rho_i=+1$) and negative ($\rho_i=-1$) scalar parts of the free
one-nucleon propagator
\begin{equation}\label{}
S^{(l)}_{\rho_i}({k_l}_0,|\mathbf
k_l|)=\dfrac{1}{E_{{k_l}}-\rho_i\Bigl(\tfrac{M_d}{2}-(-1)^l{k_l}_0\Bigr)}.
\end{equation}
and factors $\tilde\Gamma_{Sm_s\lambda m_d}^a(\omega,\mathbf
k_l;l,\alpha)$ absorb the spin-angular part of the reaction
amplitude. They also depend on the Lorentz boost parameter
$\sqrt{E_\gamma}/s^{1/4}$.

A reduce code is set up for the explicit analytical evaluation of
the spin-angular factors with free polarization indices in square
brackets of the formula~\eqref{t-final} for each isospin $a$,
particle $l$ and state $\alpha$ numbers.

\section{\label{sec:results}Results}

It finally remains to numerically calculate the
Eq.~\eqref{t-final}. After that one can proceed with studying the
effects on the differential cross section and polarization
observables due to various relativistic contributions. Beforehand
it is worthwhile to note that the BSE for the OBE interaction as
previously described is solved in the energy-momentum space using
the Wick rotation in the relative energy variable. That implies
that the deuteron vertex function is obtained along the imaginary
relative energy axis $k_4$. As seen from Eq.~\eqref{t-final} in
calculating the reaction amplitude, we are to obtain the vertex
function at some real ${k_l}_0$ axis. But it is rather hard to do,
since numerical procedure is very unstable. For not high ${k_l}_0$
values by expanding the deuteron vertex function in a Taylor
series around $k_4=0$, we can principly find the function at given
value of the relative energy variable. In numerical work we may
use the analytical parameterization of the BS amplitude for the
deuteron presented in Ref.~\cite{umnikov-param}. Unfortunately,
our examination shows that the derivatives of eight components
$g(k_4,|\mathbf k|;\alpha)$ of the vertex function with respect to
the relative energy can not be calculated with a sufficient
controlled accuracy. In Ref.~\cite{mypaper1} it was shown that a
very good approximation to the exact result is the zeroth order
approximation for the BS amplitude (BS-ZO). It comes to keeping
the $k_0$-dependence of the one-particle propagator in
Eq.~\eqref{t-final}, but keeping the relative energy ${k_0}_l$ in
the radial part $g$ of the vertex function equal to zero. That
means that both nucleons are forced to stay equally off  the
mass-shell. The Lorentz boost on the angular momentum part of the
vertex function is included as well. This approximation implies
that the retardation due to dependence of the radial part of the
vertex function on the $k_0$ is far less important than the boost
on the one-particle propagator due to recoil. It is also justified
by conclusions drawn from the studies of the elastic
electron-deuteron scattering~\cite{graz-tjon}. When ${k_l}_0=0$,
the odd in the relative energy $^1P_1^{\text{o}}$ and
$^1P_1^{\text{e}}$ components of the vertex function vanish. We
may hope that this is a minor drawback, since the net
presudoprobability of these states $P_{\text{odd}}$ is roughly one
order of magnitude less than that of the two even $P$-states
(compared with the value of $P_{\text{even}}$ in
Table~\ref{tab:table1}). As for the separable interaction we can
only do exact calculations of the deuteron vertex function in the
$^3S_1^+$--$^3D_1^+$ channel at a given value of the relative
energy variable. The negative-energy states are completely left
behind in this case.

\begin{figure}[ht]
\centering
\includegraphics[width=0.75\textwidth]{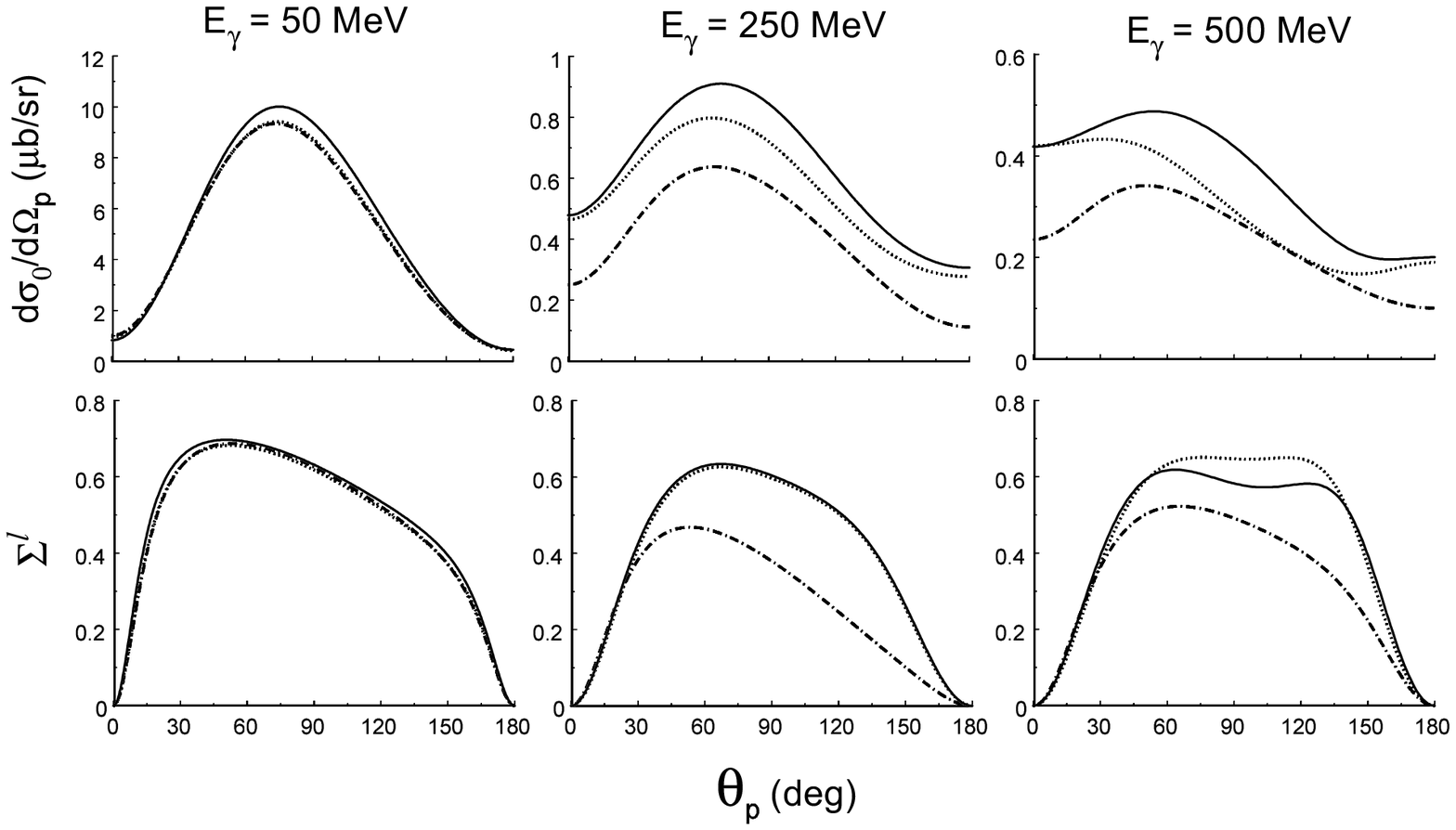}
\caption{\label{fig2} The differential cross section and the
linear photon asymmetry in the plane wave one-body approximation
at different lab. photon energies $E_\gamma$. Curves: solid line,
the OBE interaction vertex function computed in the BS-ZO
approximation (6 partial-wave states are included); dotted line,
the OBE interactions with taking into account only two
positive-energy partial-wave states of the deuteron vertex
function; dot-dashed line, calculation with the deuteron vertex
function in case of the separable interactions.}
\end{figure}

In Figs.~\ref{fig2} the results for the angular distributions of
the differential cross section $d\sigma_0/d\Omega_p$ and the
linear photon asymmetry $\Sigma^l$ at three different values of
lab photon energies are presented.
The BS-ZO calculation in the OBE model is given by the solid
curve. The dotted line shows the effect of shutting off all the
negative energy states of the OBE vertex function of the deuteron.
That leaves only two positive energy $L=0$ and $L=2$ states. The
dot-dashed curve should be compared with the dotted one. The
former depicts the calculation in the separable model with the
inclusion of all the retardation factors and relativistic effects
apart from those generated by the negative energy partial waves.
As can be seen from Fig.~\ref{fig2}, the global structure of the
observable is defined by the positive energy components of the
deuteron vertex function. The negative energy components increase
markedly the cross section but leave its global structure
untouched. Comparing the dotted and dot-dashed line, we observe
the systematic deficiency of the calculations with the separable
positive energy states of the vertex function with respect to the
OBE one. Explanations can be found observing Fig.~\ref{fig1}. As
follows form the Eq.~\eqref{diff} the differential cross section
is proportional to square of the modulus of the partial-wave
components of the vertex function multiplied by the over-all
kinematic factor. At $E_\gamma=50$~MeV the absolute value of the
three relative momentum of the $l$th bound nucleon $|\mathbf
k_l|$, see the Eq.~\eqref{kl}, lies in the range 200--265~MeV. At
this momentum range the separable and OBE vertex function are
almost identical and, thus, both functions give the same result.
The tiny rise of the differential cross section in the case of the
OBE interaction is due to the smallness of the negative energy
components of the deuteron vertex function. Results start to be
closely interaction kernel dependent as the photon energy rises.
At $E_\gamma=500$~MeV the absolute value of $|\mathbf k_l|$ is
within the interval 450--950~MeV. As seen in Fig.~\ref{fig1} the
$^3S_1^{+}$-partial wave in case of the OBE interactions has a
more pronounced minimum than that of the Graz-II interaction
kernel. We can conclude that the cross section is rather sensitive
to the momentum behavior of the partial-wave components of the
vertex function. As for the negative energy components the
$^1P_1^{\text{e}}$- and $^3P_1^{\text{o}}$-states have a
persistent superiority over the positive-energy states at high
three momentum. As the contribution of the latter is diminished,
the former increases the cross section. Effects produced
$^3S_1^{-}$- and $^3D_1^{-}$-states are completely negligible.
Shutting off the negative energy states has a minor effects,
particularly, on the photon asymmetry in a wide photon energy
range. Though the reaction amplitude is subject to a change when
the relativistic components of the vertex function are included,
we are to suppose that they cancel each other in the photon
asymmetry.

\begin{figure}[b]
\centering
\includegraphics[width=0.75\textwidth]{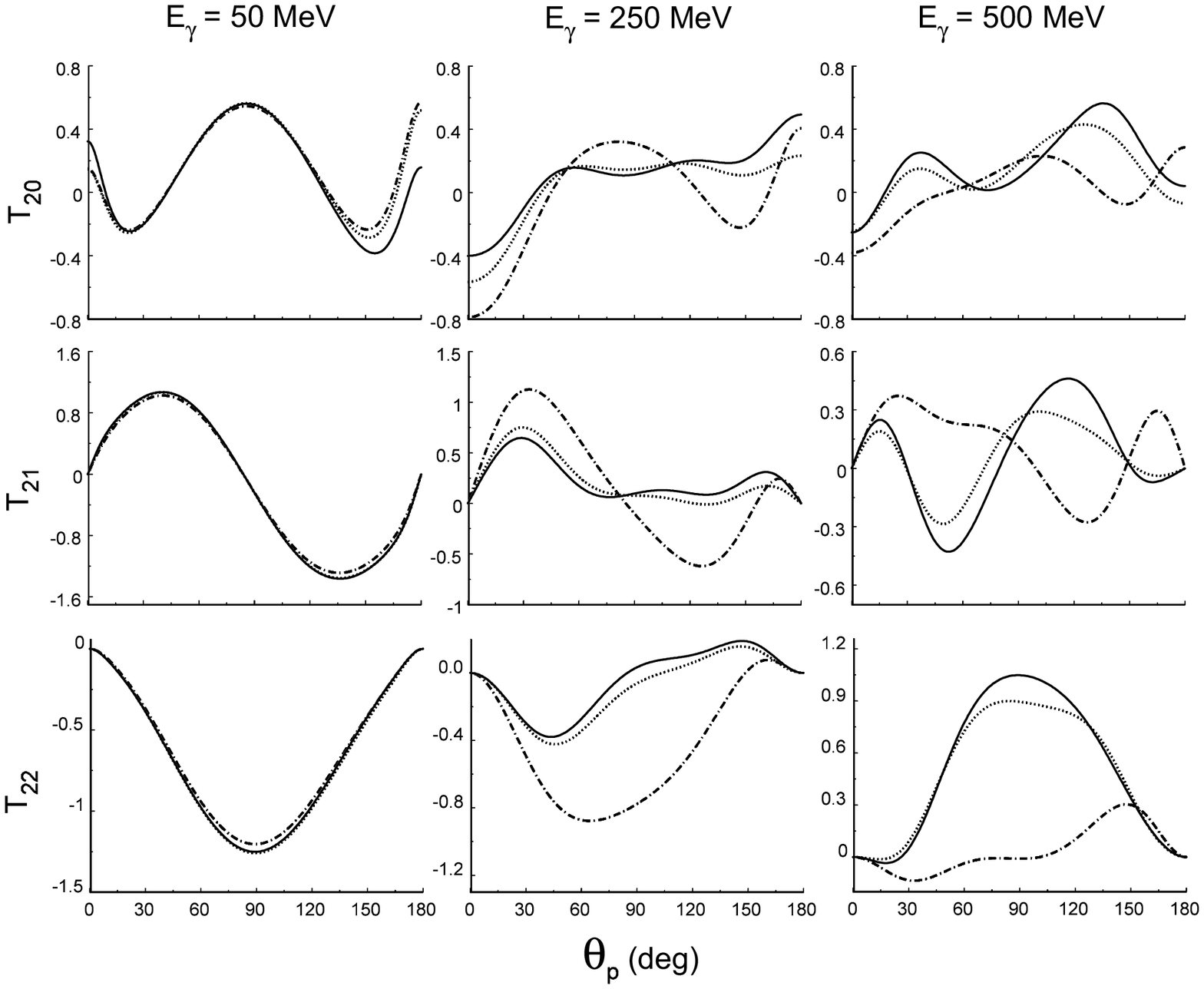}
\caption{\label{fig3} The tensor target asymmetries. Notations are
the same as in Fig.~\protect{\ref{fig2}}. }
\end{figure}

Next, we discuss three observables associated with tensor
polarization of the deuteron. Fig.~\ref{fig3} shows the angular
dependence of the tensor target asymmetries.
The solid curve is the calculation with the six partial wave
components of the OBE vertex function of deuteron. Comparison with
the dot curve shows what differences arise when the negative
energy states are invoked to describe the physical process.
Inclusion of these states at a moderate photon energy effects to a
change of the $T_{2M}$ at the forward and backward proton c.~m.
angles $\Theta_p$. At higher energies these are visible in the
whole $\Theta_p$ range. As in the case of the differential cross
section, the negative energy states leave the global structure
inalterable. Again, the dot-dashed line should be compared with
the dotted one. For the photon energies about 500 GeV two curves
demonstrate sensitivity bearing by the observebles on the model of
the NN interaction. The dissimilarity between the OBE and
separable models clearly shows up in the observable $T_{22}$. For
$M=2$ in the Eq.~\eqref{t2m} there are not interference terms
between amplitudes $t_{Sm_s \lambda m_d}$ with different $m_d$.
The Clebsch-Gordan coefficient in the Eq.~\ref{t2m} for the
observable $T_{22}$ selects reaction amplitudes $t_{Sm_s \lambda=1
m_d}$ with $m_d=-1$. In contrast with that the $T_{20}$ is the sum
of products of reaction amplitudes with all allowed values of
$m_d$.

\section{\label{sec:remarks}Concluding remarks}

In the previous sections we presented the results of the
relativistic covariant calculation of the polarization
observables in two-body deuteron photodisintegration. Our aim has
been to study contribution of the relativistic components of the
deuteron amplitude, which does not have counterparts in the
nonrelativistic physics. These negative energy components comply
with requirements of covariance, since the full Dirac structure
should be taken into account in the partial-wave analysis of the
full BS amplitude of the NN system.

In this paper we have studied the influence of the relativistic
effects on the photon beam and three tensor target asymmetries as
well as the differential cross section within the framework of the
BS formalism. The significance of the negative energy components
is examined using the BS equation for the OBE interaction kernel,
which couples two positive and six negative energy states of the
deuteron vertex function. Moreover, the importance in
particularities of the behavior of the positive energy components
is tested as well. We compare the results obtained for the
observables in the OBE model with those of the multirank separable
model of the NN interaction; in the separable model the negative
energy states are switched off.

Unfortunately, our results do not account properly for the role of
the relative energy variable $k_0$ of the deuteron vertex
function. A fundamental obstacle here is posed via the numerical
treating of the BSE. After the partial-wave decomposition of the
equation, the numerical procedure in solving the eigenvalue
problem is based on the Wick rotation. The components of the
vertex function are computed along the imaginary $k_4=\imath k_0$
axis. This technique prevents a direct application and practical
use of the vertex function in some physical process involving
deuteron. Considering reliable possibilities of making approximate
calculations, we employ the zeroth order approximation for the
radial part of the vertex function. It allows one to take into
account six out of eight partial wave states as well as other very
important relativistic effects.

In studying the cross section and asymmetries, we establish the
following results. Adding the negative energy states of the OBE
vertex function to the positive energy, we find lead to a sizeable
increasing (up to 10~\%) of the differential cross section in a
wide range of the proton c.~m angles. This tendency becomes more
pronounced at higher photon energies. The asymmetries are less
influenced by these states and, thus, show minor changes. However,
modifications become noticeable in the tensor asymmetries at the
deuteron break-up in the forward and backward directions. On the
other hand, the results indicate strong dependence on the behavior
of the $^3S_1^{+}$--$^3D_1^+$-partial waves as functions of the
relative three momentum of nucleons.

Numerical results of this paper has been obtained in the
plane-wave approximation with one-body EM current operator. Since
the one-body current is not conserved, the choice of a gauge for
the radiation field does become important. We have used the
transverse gauge, that in combination with the one-body current
leads to a too small value for the matrix elements. The two-body
nucleonic current as well as exchange currents should be called
into play to restore the gauge independence of the reaction
amplitude and the Siegert limit (see, for example,
Refs.~\cite{benz,ying}). Therefore, an inclusion of the two-body
EM currents into the relativistic analysis is of the prime
importance. Corresponding numerical work will be done in future
and results are going to be reported in a forthcoming paper.

\begin{acknowledgments}

We would like to thank A.A.~Goy and S.Eh.~Schirmovsky  for helpful
conversations. We also wish to thank S.M.~Dorkin and L.P.~Kaptari
for consultations in conducting the numerical part of this work.
We are grateful to O.V.~Chelomina for the assistance rendered in
preparation of the paper. This research was supported in part by
Grant No.~015.02.01.022, "Universities of Russia".
\end{acknowledgments}

\clearpage\newpage

\end{document}